\documentclass[twocolumn]{aastex631}

\usepackage[english]{babel}
\usepackage{amsmath}
\usepackage{amssymb}
\usepackage{graphicx}
\usepackage{subfigure}
\usepackage[normalem]{ulem}
\usepackage{enumerate}
\usepackage{booktabs}

\usepackage{verbatim}

\usepackage{color}
\usepackage{multirow}
\usepackage{mathtools}
\usepackage{epstopdf}
\usepackage{tabularx}

\usepackage{url}
\usepackage{xcolor}

\def\nat{Nature}

\def\mnras{MNRAS}
\def\apj{ApJ}
\def\apjl{ApJL}
\def\apjs{ApJS}
\def\aap{A\&A}

\def\araa{ARA\&A}

\def\pasj{Pub. Astron. Soc. Japan}

\def\be{\begin{equation}} 
\def\ee{\end{equation}} 
\def\ba{\begin{eqnarray}} 
\def\ea{\end{eqnarray}}

\def\gsim{\lower.5ex\hbox{\gtsima}} 
\def\lsim{\lower.5ex\hbox{\ltsima}} \def\gtsima{$\; \buildrel > \over 
\sim \;$} \def\ltsima{$\; \buildrel < \over \sim \;$} \def\prosima{$\; 
\buildrel \propto \over \sim \;$} \def\gsim{\lower.5ex\hbox{\gtsima}} 
\def\lsim{\lower.5ex\hbox{\ltsima}} 
\def\simgt{\lower.5ex\hbox{\gtsima}} 
\def\simlt{\lower.5ex\hbox{\ltsima}} 
\def\simpr{\lower.5ex\hbox{\prosima}}   
  
 \def\gtsima{$\; \buildrel > \over \sim \;$} 
\def\ltsima{$\; \buildrel < \over \sim \;$} 
\def\gsim{\lower.5ex\hbox{\gtsima}} 
\def\lsim{\lower.5ex\hbox{\ltsima}} 
\def\simgt{\lower.5ex\hbox{\gtsima}} 
\def\simlt{\lower.5ex\hbox{\ltsima}} 
\def\simpr{\lower.5ex\hbox{\prosima}}

\def\E3{{\cal E}_{\rm g}^{III}}

\def\r12{r_{1/2}} 
\def\x12{x_{1/2}} 
\def\v12{v_{1/2}}

\def\nh2{n_{\rm H2}}
\def\fh2{f_{\rm H2}}

\def\angstrom{\textrm{A\kern -1.3ex\raisebox{0.6ex}{$^\circ$}}}

\makeatletter
\def\@hex@@Hex#1%
 {\if a#1A\else \if b#1B\else \if c#1C\else \if d#1D\else
  \if e#1E\else \if f#1F\else #1\fi\fi\fi\fi\fi\fi \@hex@Hex}
\makeatother

\definecolor{apcolor}{HTML}{b3003b}
\definecolor{cbcolor}{HTML}{ff0f00}
\definecolor{afcolor}{HTML}{b3443c}
\definecolor{vgcolor}{HTML}{8F00FF}
\definecolor{tbdcolor}{HTML}{E8A95E}
\definecolor{stefcolor}{HTML}{0047ab}

\newcommand\dodoi[1]{doi:~\href{http://doi.org/#1}{\nolinkurl{#1}}}

\newcommand\doarXiv[1]{\href{https://arxiv.org/abs/#1}{\nolinkurl{https://arxiv.org/abs/#1}}}

\graphicspath{{./}{figures/}}

\submitjournal{ApJL}

\shorttitle{True PISN discendant: Pop~III IMF implications}
\shortauthors{Koutsouridou, Salvadori et al.}

\begin{document}

\title{True Pair-instability Supernova Descendant: Implications for the First Stars’ Mass Distribution}

\correspondingauthor{Ioanna Koutsouridou}
\email{ioanna.koutsouridou@unifi.it}

\author[0000-0002-3524-7172]{Ioanna Koutsouridou}
\author[0000-0001-7298-2478]{Stefania Salvadori}
\author[0000-0001-9155-9018]{\'{A}sa Sk\'{u}lad\'{o}ttir}
\affiliation{Dipartimento di Fisica e Astronomia, Universit\'{a} degli Studi di Firenze, Largo E. Fermi 1, 50125, Firenze, Italy}
\affiliation{INAF/Osservatorio Astrofisico di Arcetri, Largo E. Fermi 5, 50125, Firenze, Italy}

\begin{abstract}
The initial mass function (IMF) of the first Pop~III stars remains a persistent mystery. Their predicted massive nature implies the existence of stars exploding as pair-instability supernovae (PISN), but no observational evidence had been found. Now, the LAMOST survey claims to have discovered a pure PISN descendant, J1010+2358, at ${\rm[Fe/H]}=-2.4$. Here we confirm that a massive 250-260$\:{\rm M_\odot}$ PISN is needed to reproduce the abundance pattern of J1010+2358. However, the PISN contribution can be as low as 10$\%$, since key elements are missing to discriminate between scenarios. We investigate the implications of this discovery for the Pop~III IMF, by statistical comparison with the predictions of our cosmological galaxy formation model, {\sc NEFERTITI}. First, we show that the non-detection of mono-enriched PISN descendants at ${\rm [Fe/H]}<-2.5$ allows us to exclude: (i)~a flat IMF at a 90$\%$ confidence level; and (ii)~a Larson type IMF with characteristic mass $m_{\rm ch}/{\rm M_\odot}>191.16x-132.44$, where $x$ is the slope, at a 75$\%$ confidence level. Secondly, we show that if J1010+2358 has only inherited $<70\%$ of its metals from a massive PISN, no further constraints can be put on the Pop~III IMF. If, instead, J1010+2358 will be confirmed to be a nearly pure ($>90\%$) PISN descendant, it will offer strong and complementary constraints on the Pop~III IMF, excluding the steepest and bottom-heaviest IMFs: $m_{\rm ch}/{\rm M_\odot}<143.21x-225.94$. Our work shows that even a single detection of a pure PISN descendant can be crucial to our understanding of the mass distribution of the first stars. 

\end{abstract}


\section{Introduction}
\label{sec:intro}

The first (Pop~III) stars formed out of primordial composition gas and thus were likely more massive than those observed today, with a mass range possibly extending up to $\approx 1000 \:{\rm M_\odot}$ \citep[e.g.][]{Hirano2015}. These theoretical findings have strong physical grounds and have been supported across decades by both analytical calculations \citep[e.g.][]{McKeeTan2008} and numerical simulations \citep[e.g.][]{Susa2014}. Furthermore, in recent years, stellar archaeology has provided novel data-driven constraints on the properties of the first stars, which confirm their massive nature \citep[e.g.][]{Hartwig2018a,rossi2021ultra,Koutsouridou2023}. However, the initial mass function (IMF) of the first stars is largely unknown and it is still unclear if primordial stars with hundreds of solar masses were really able to form \citep[e.g.][for a recent review]{Klessen2019}. 

Very massive first stars, $140 \:{\rm M_\odot} \leq m_\star \leq 260 \:{\rm M_\odot}$, are predicted to end their lives as energetic Pair Instability Supernovae (PISN), which completely destroy the progenitor star, injecting into the interstellar medium (ISM) $\approx 50\%$ of their mass in the form of heavy elements. Stellar evolution calculations for PISNe are very robust (Limongi, private comm.) and predict the gas to be imprinted with a {\it unique} chemical signature, showing a strong odd-even effect \citep{Heger2002, Takahashi2018}. 
PISN descendants, i.e. long-lived stars formed in gaseous environments predominantly imprinted by PISNe, can potentially be found in the most ancient component of our Milky Way (MW) and its dwarf galaxy satellites. They are, however, predicted to be extremely rare, with a metallicity distribution function peaking at $\mathrm{[Fe/H] \approx-2.0}$ \citep[e.g.][]{karlsson2008, deBen2017}.

After decades of dedicated searches \citep[e.g.][]{Beers2005,Aoki2014,caffau2023} and development of novel techniques to pin-point PISN descendants \citep{salvadori2019probing, Aguado2023}, in June 2023 a remarkable Galactic halo star with very strong odd-even effects has finally been identified by the LAMOST survey \citep[][]{Xing2023}. This very metal-poor star, J1010+2358, has $\rm[Fe/H]\approx -2.42 \pm 0.12$, and exhibits extremely low abundances of odd elements such as Na and Sc. Furthermore, it shows very large abundance variance between the odd and even elements, such as [Na/Mg] and [Co/Ni]. Thus, \citet{Xing2023} claimed it to be a ``pure descendant" of a very energetic PISN progenitor with a mass of $260 \:{\rm M_\odot}$.

Ultimately, despite the identification of stars that might have been {\it partially} imprinted by PISNe \citep{Aoki2014,salvadori2019probing,caffau2023,Aguado2023}, J1010+2358 represents so far the unique candidate of a {\it pure} PISN descendant. But is the abundance pattern of J1010+2358 only consistent with {\it mono-enrichment} by a single $260 \:{\rm M_\odot}$ PISN, or do other solutions exist? What are the implications of the detection of a {\it single} PISN descendant with ${\rm[Fe/H]}\approx -2.5$ for the IMF of the first stars? The aim of this Letter is to address these questions by comparing the observed frequency of PISN descendants with predictions from state-of-the-art cosmological models for the MW formation. 

\section{Observations: PISN descendants}
\label{Data}

To discriminate among model predictions which assume different Pop~III IMFs, we should quantify the observed frequency of PISN descendants. However, we first need to clarify whether J1010+2358 has truly been mono-enriched by a single $260 \:{\rm M_\odot}$ PISN or if other enrichment channels are also possible.

\begin{figure*}
    \includegraphics[width=0.55\hsize]{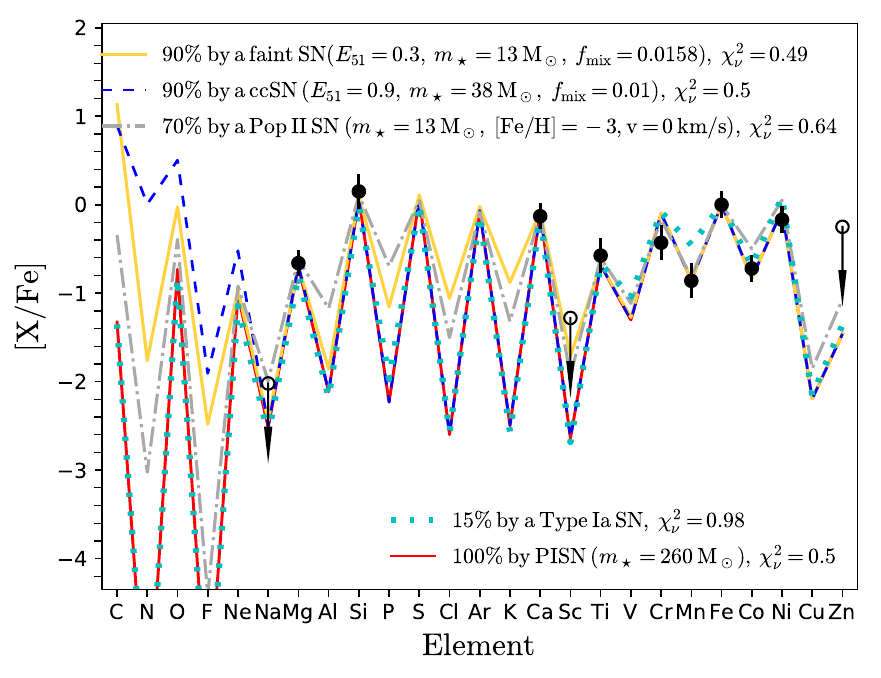} 
    \includegraphics[width=0.45\hsize]{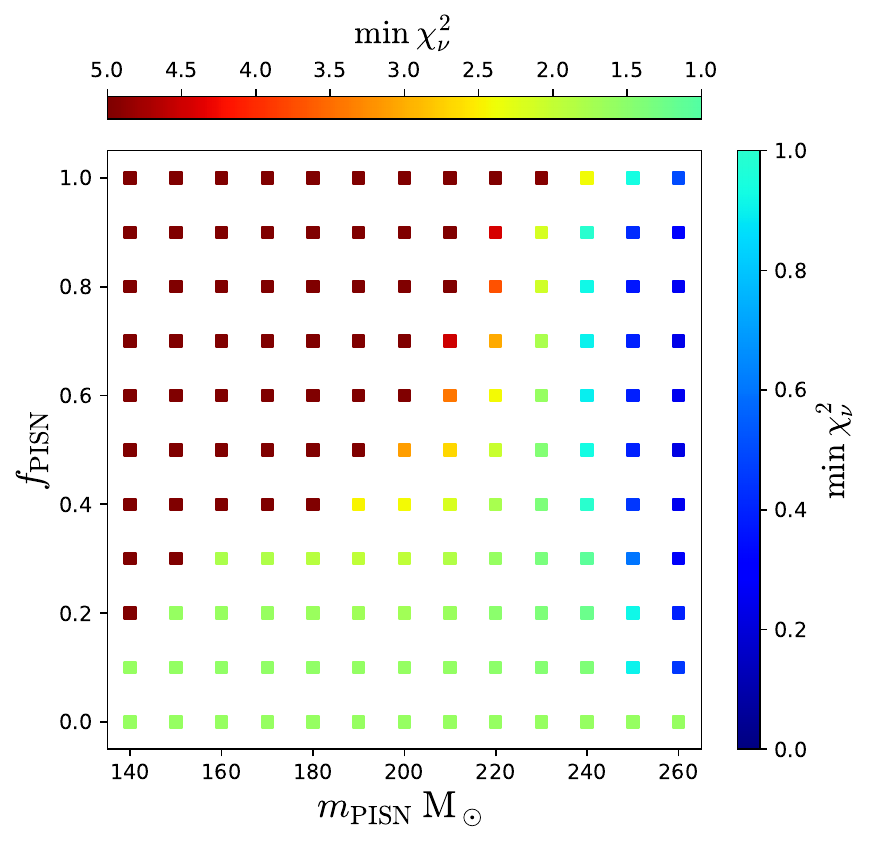} 
    \caption{{\it Left:} measured chemical abundances of J1010+2358 (black points with errorbars) compared to: (i) a $260\:{\rm M_\odot}$ PISN (red); (ii) 10$\%$ of metals by a $260\:{\rm M_\odot}$ PISN and 90$\%$ by a primordial faint SN (yellow) or a primordial core-collapse SN (blue; yields by \citealp{Heger2010}); (iii) 30$\%$ of metals by a massive PISN and 70$\%$ by a Pop II SN (gray; \citealp{Limongi2018}); (iv) 85$\%$ of metals by a massive PISN and 15$\%$ by a Type Ia SN (cyan; model W70 by \citealp{Iwamoto1999}). {\it Right:} minimum $\chi_\nu^2$ of the fit to J1010+2358's chemical abundances with different fractions of PISN enrichment, $f_{\rm PISN}$, from a mass $m_{\rm PISN}$ progenitor, and the rest from Pop~III SNe.}
  \label{fig:Xing_fit}
\end{figure*}

\subsection{Is J1010+2358 mono-enriched by a PISN?}
\label{data-monoenrich}

In Fig.~\ref{fig:Xing_fit} (left) we see that the observed abundance pattern\footnote{For Ti, we use the average of the Ti I and Ti II measurements, which have a difference of 0.15 dex.} of J1010+2358 is in perfect agreement with the one predicted for a pure enrichment by a $260 \:{\rm M_\odot}$ PISN, as stated by \citet{Xing2023}. The fit of the model to the observed data is extremely good, $\chi_\nu^2 = \chi^2/\nu = 0.5$, where $\nu$ is the total number of observed data points, including normal ($N$) measurements, upper ($U$) and lower ($L$) limits, and: 
\begin{equation}
    \chi^2 = \sum_{i=1}^N  \frac{(F_i-D_i)^2}{\sigma^2_i} + \sum_{i=N+1}^{N+U+L}  \frac{(F_i-D_i)^2}{\sigma^2_i} \Theta (F_i - D_i),
\label{e:chi}
\end{equation}
following \cite{Heger2010}. Here, $F_i$ and $D_i$ are the theoretical and observed values of [X/Fe], respectively, $\sigma_i$ are the observational uncertainties, and $\Theta(x)=1$ for $x > 0$ ($x < 0$) if $D_i$ is an upper (lower) limit, and $x=0$ otherwise.

To understand whether J1010+2358 could also be ``multi-enriched", we analyze the simplest case, that is an enrichment by two different SNe.
We assume that a fraction $f_{\rm PISN}$ of its metals has been contributed by a PISN with mass $m_{\rm PISN}=[140-260]\:{\rm M_\odot}$, while the remainder comes from a second SN, which can be of any other type: a Pop~III SN (ranging in energy, mixing, and mass) or a Pop~II SN (ranging in mass and metallicity), thus more generally named ${\rm SN}_j$. The chemical abundance pattern of a star imprinted by these two SNe would be:
\begin{multline}
     F_i = {\rm [X/Fe]} = {\rm log} \left( \frac{{\rm Y_X^{PISN}}+\beta\frac{\rm Y_Z^{PISN}}{{\rm Y_Z}^{{\rm SN}_j}}{{\rm Y_X}^{{\rm SN}_j}}}{{\rm Y_{Fe}^{PISN}}+\beta\frac{\rm Y_Z^{PISN}}{{\rm Y_Z}^{{\rm SN}_j}}{{\rm Y_{Fe}}^{{\rm SN}_j}}} \right) + \\
    - {\rm log} \left( \frac{\rm M_X}{\rm M_{Fe}} \right)_\odot   
\end{multline}
\citep{salvadori2019probing}, where $\beta = (1-f_{\rm PISN})/f_{\rm PISN}$, ${\rm Y_X}^{{\rm PISN}}$, ${\rm Y_X}^{{\rm SN}_j}$ are the theoretical elemental yields and ${\rm Y_Z}^{{\rm PISN}}$, ${\rm Y_Z}^{{\rm SN}_j}$ the total metal yields of the PISN and ${\rm SN}_j$, respectively (see the caption of Fig.~\ref{fig:Xing_fit} for the adopted yields). 
By substituting the above in Eq.~\ref{e:chi} we determine the best fit over all $\rm SN_j$ (minimum $\chi_\nu^2$) to the abundance pattern of J1010+2358, for each {$f_{\rm PISN}$ and} $m_{\rm PISN}$. 

The results are shown in Fig.~\ref{fig:Xing_fit} (right).
We find that a contribution from a PISN with $250-260 \:{\rm M_\odot}$ is necessary to reproduce the abundance pattern of J1010+2358: $\chi_\nu^2<1$ for $m_{\rm PISN}\geq250\:{\rm M_\odot}$ and $f_{\rm PISN}\geq 0.1$. 
The goodness of fit in each case depends on the properties of the secondary SN. Even for $f_{\rm PISN}=0.1$, there exist numerous Pop~III $\rm SN$ of different type $j$ that can yield $\chi_\nu^2\leq1$. Two such examples are shown in the left panel of Fig.~\ref{fig:Xing_fit}. When Pop II SNe are investigated, we find that a minimum contribution of $f_{\rm PISN}=0.3$ is required to match J1010+2358's abundance pattern (Fig.~\ref{fig:Xing_fit}, gray line). As in the case of Pop~III SNe, the massive PISN contribution here is necessary to reproduce the observed strong odd-even effect or/and the low [Na/Fe] and [Sc/Fe] ratios. Furthermore, $\chi_\nu^2<1$ only when the contribution from Type Ia SNe is $<15\%$ (Fig.~\ref{fig:Xing_fit}, cyan line). Otherwise, the iron-peak elements Mn, Ni and Co are overrepresented against the lighter elements Mg and Si. Nevertheless, the contribution of Type Ia SNe to second-generation stars is likely minimal due to their typical delay times \citep{Komiya2016}.

These results indicate that J1010+2358 could be imprinted by a $250-260\:{\rm M_\odot}$ PISN and one or multiple SNe. Alternatively, there exists the possibility that J1010+2358 has been completely enriched by peculiar 12-14$\:{\rm M_\odot}$ ccSNe that experience negligible fallback \citep{Jeena2023}.

\subsection{Frequency of mono-enriched PISN descendants}
\label{data-fraction}
The star J1010+2358 is very metal-poor (VMP; $\rm [Fe/H]\leq-2$) with sub-solar [Mg/Fe] based on the low-resolution LAMOST survey \citep{Zhao2012}, which comprises over 15\,000 VMP stars \citep{Li2022,Aoki2022}. J1010+2358 was followed up at high resolution and confirmed as a probable PISN descendant \citep{Xing2023}.
Since only a small subset of the LAMOST sample has been followed up at high resolution, more of these stars could possibly harbor an imprint from a 250-260$\:{\rm M_\odot}$ PISN. Therefore, we can treat the identified fraction of massive PISN descendants, $\sim$1/15\,000, as a lower limit to compare with our theoretical predictions (Fig.~\ref{fig:mch}, bottom). 

Using the SAGA\footnote{\url{http://sagadatabase.jp/}} database \citep[e.g.][]{Suda2008, Suda2017}, 
we examine whether more VMP stars have abundance patterns consistent with mono-enrichment by a PISN. Searching for a partial PISN enrichment would be inconclusive, since contribution from other SNe can mask the distinctive PISNe signature. We restrict our analysis to Galactic halo stars with at least six elemental abundance measurements (excluding upper limits), and excluding CEMP-s stars, since their abundance patters are not representative of their birth environments. In case of multiple entries from different surveys/authors for the same star, we keep the one with the most measured abundances. We end up with a sample of 962 unique stars with $-5\leq{\rm [Fe/H]}< -2.5$. 

Most chemical abundances in our sample have been measured assuming a one-dimensional, local thermodynamic equilibrium (LTE) hydrostatic model atmosphere. Following \citet{Ishigaki2018}, we thus assign a large observational uncertainty of 0.3~dex to the abundances of C, N, O, Na and Al that are strongly affected by non-LTE effects. Additionally, we adopt a 0.2~dex error for Si, Ti, Cr and Mn because their measured abundances vary significantly depending on the absorption lines used to derive them. For all other elements, we assume a 0.15~dex error, unless a larger uncertainty is reported in the literature. 
Due to the large errors assumed, we require that $\chi_\nu^2\leq 1$ for a fit to be considered good. 

As expected, zero stars from the SAGA sample at ${\rm [Fe/H]}<-2.5$ are consistent with enrichment by a single PISN progenitor. We use this non-detection to adopt upper limits on the fraction of mono-enriched PISN descendants (Fig.~\ref{fig:mch}, top).

\section{Theory: summary of {\sc NEFERTITI}}
\label{sec:model}

The code {\sc NEFERTITI} (NEar FiEld cosmology: Re-Tracing Invisible TImes; details in \citealt{Koutsouridou2023}), is a state-of-the-art cosmological chemical evolution model intended to study the unknown properties of the first stars (IMF and energy distribution of the first SNe).
{\sc NEFERTITI} can run on halo merger trees obtained from N-body cosmological simulations or Monte-Carlo techniques, and grounds on previous semi-analytical models for the Local Group formation \cite[][]{Salvadori2007, Salvadori2015, Pagnini2023}. Here we employ {\sc NEFERTITI} coupled with a cold dark matter (DM) N-body simulation of a MW analogue (fully described in \citealt{Koutsouridou2023}), which succesfully reproduces the present-day global properties of the MW (metallicity and mass of both stars and gas) and the metallicity distribution function of the Galactic halo \citep{Bonifacio2021}. In the following, we recap the main assumptions and innovative aspects of the semi-analytical model.

{\sc NEFERTITI} follows the evolution of the baryonic component within DM halos based on the following assumptions: 
(i)~at the highest redshift of the simulation the intergalactic medium (IGM) has a primordial composition; (ii)~gas from the IGM is continuously accreted onto each halo at a rate proportional to the halo's DM growth and subsequently streams onto the halo's central galaxy in freefall; (iii)~stars form in DM halos that exceed a minimum mass, which evolves through cosmic times to account for the photo-dissociating and ionizing radiation \citep{salvadori2009}; (iv)~within each galaxy, stars form on a rate proportional to the available gas mass, the rate being reduced in minihalos (with virial temperature $T_{\rm vir}\leq 2 \times 10^4\:$K) to account for the ineffective cooling by molecular hydrogen \citep{salvadori2012}; (v)~gas and metals that are returned through stellar winds and SNe to the ISM and the IGM, are assumed to be instantaneously mixed; (vi)~at each timestep, the stellar, gas and metal masses within each DM halo are equally distributed among its particles and remain attached to them to the next integration step.\\

The most innovative aspects of {\sc NEFERTITI} are that the code accounts for the:
\begin{itemize}
    \item incomplete sampling of the stellar IMF \citep{rossi2021ultra} for Pop~III and Pop~II/I stars;
    \item unknown Pop~III IMF, parameterized as \citep{larson1998early}:     
    \begin{equation}
    \phi(m_\star) = \frac{d N}{d m_\star} \propto m_\star^{-x} {\rm exp} \bigg( - \frac{m_{\rm ch}}{m_\star} \bigg) 
     \label{e:IMF}
    \end{equation}
    where $m_\star = [0.8-1000]\:{\rm M_\odot}$ following \citet{rossi2021ultra}, and the characteristic mass, $m_{\rm ch}$, and IMF slope, $x$, are varied; while normal Pop~II/I stars form according to a Larson IMF with $m_{\rm ch}=0.35\:{\rm M_\odot}$, $x=2.35$ and $m_\star = [0.1-100]\:{\rm M_\odot}$ when the ISM has a metallicity $Z \geq Z_{\rm crit} = 10^{-4.5}\:{\rm Z_\odot}$ \citep{deBen2017};
    \item evolution of individual stars in their proper timescales, adopting the yields of \citet{Heger2002,Heger2010} for Pop~III stars and \citet{van1997new, Limongi2018} for Pop~II/I stars;
    \item unknown energy distribution function of the Pop~III SNe ($m_\star = [10-100]\:{\rm M_\odot}$), parametrized as ${\rm d}N/{\rm d}E\propto E_\star^{-\alpha_e}$, where $\alpha_e$ can be varied. Here we adopt $\alpha_e=2$. 
\end{itemize}

In star-forming mini-halos, the available star-forming gas is sometimes less than the maximum stellar mass permitted by our assumed IMF, prohibiting the formation of massive stars and skewing the effective IMF towards low masses. 
Here, we use the following approach to adhere to our assumed IMF without favouring multi-enrichment. Whenever the random mass generator calls for the formation of a star whose mass exceeds the available star-forming gas mass, we postpone star formation in that halo until enough gas accumulates to allow that massive star's formation.

\section{Results} 
\label{results}

\begin{figure*}
\begin{center}
    \includegraphics[width=0.49\hsize]{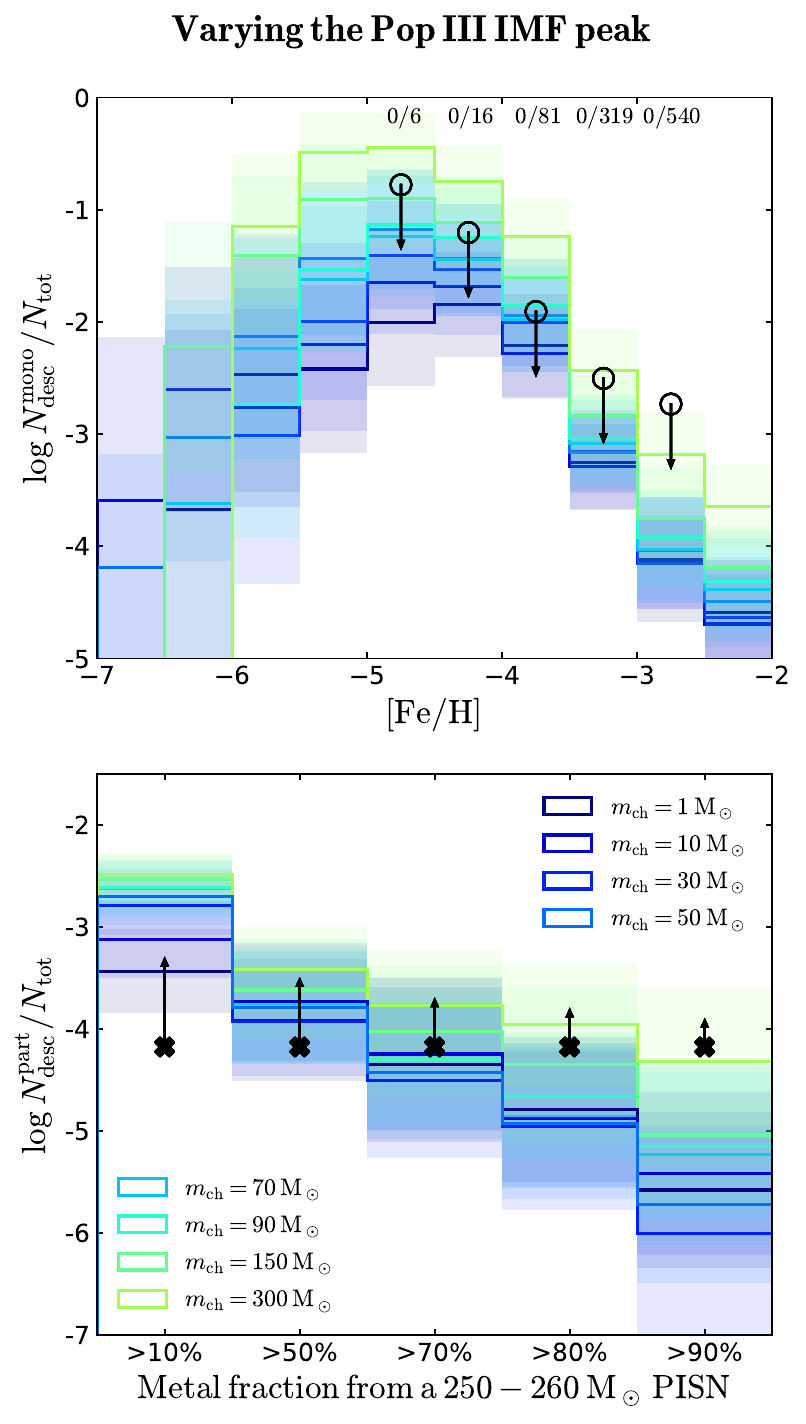} 
    \includegraphics[width=0.49\hsize]{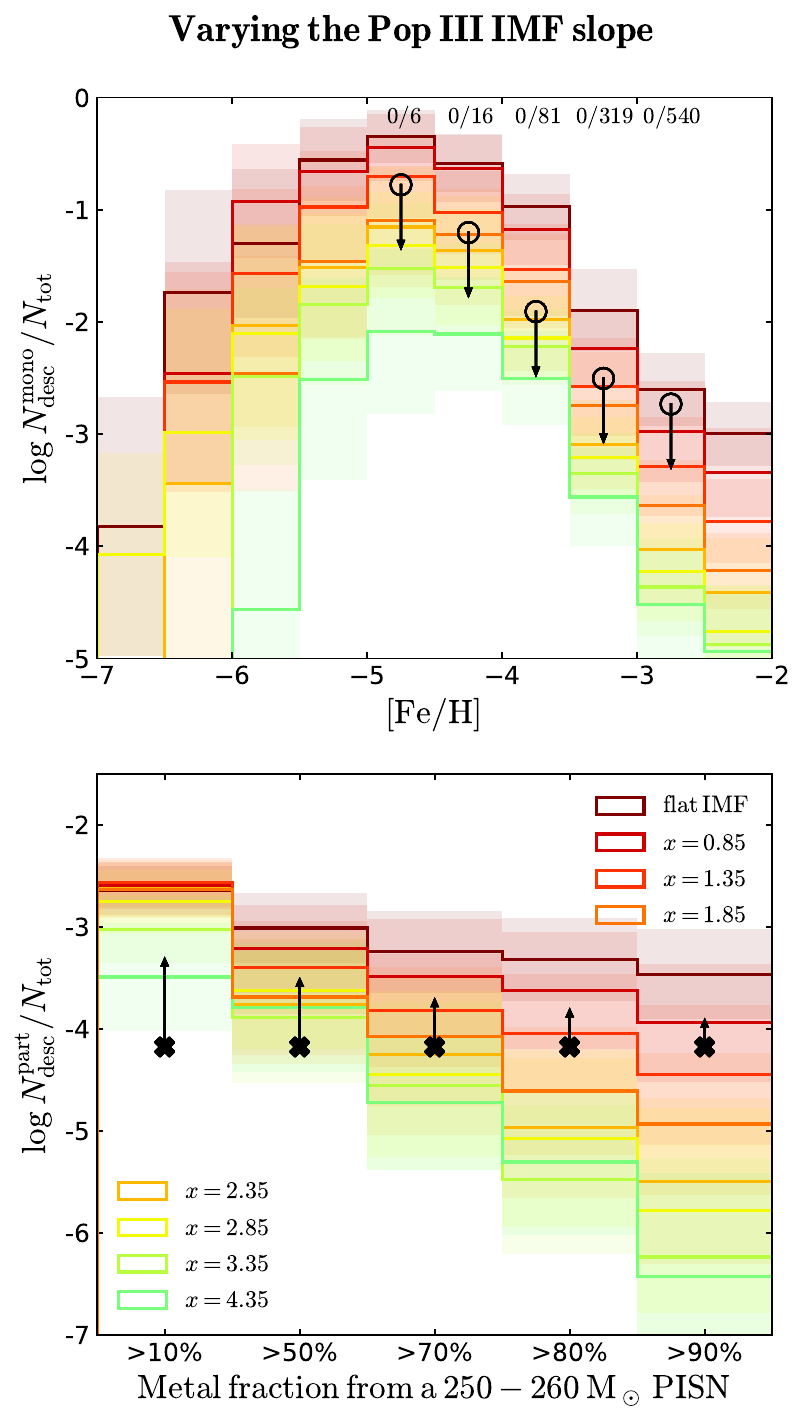} 
\caption{{\it Top}: Predicted fraction of mono-enriched PISN descendants in the inner halo as a function of [Fe/H], when varying the characteristic mass of the Pop~III IMF, for $x=2.35$ (left); and when varying $x$, for $m_{ch}=70 \:{\rm M_\odot}$ (right; Eq.~\ref{e:IMF}). A Gaussian error of 0.2\,dex has been applied to model predictions to account for observational uncertainties. Data points show the observed fractions (Sec.~\ref{data-fraction}), also listed on top. {\it Bottom}: Predicted fraction of all $\rm [Fe/H]\leq-2$ inner halo stars that have inherited at least 10, 50, 80, or 90$\%$ of their metals from a $250-260 \:{\rm M_\odot}$ PISN, for different $m_{ch}$ (left) and slope $x$ (right) of the Pop~III IMF. The X mark denotes that {\it at least} one star, J1010+2358, out of 15\,000 VMP stars in the LAMOST sample has been imprinted by a 250-260$\:{\rm M_\odot}$ PISN (Sec.~\ref{data-monoenrich}). Histograms and shaded areas represent the mean and standard deviation of 200 runs of each model. }
\label{fig:mch}
\end{center}
\end{figure*}

We can now try to constrain the Pop~III IMF through two key comparisons: (i) the fraction of mono-enriched PISN descendants with ${\rm [Fe/H]}<-2.5$ predicted to lie in the inner Galactic halo (Galactocentric radii $7\: {\rm kpc}\leq R_{\rm gal}\leq 20 \: {\rm kpc}$) at $z=0$, contrasting this with their absence in the SAGA database (Fig.~\ref{fig:mch}, top); (ii) the predicted fraction of VMP stars that have been enriched by a massive 250-260$\:{\rm M_\odot}$ PISN progenitor, in relation to the unique star so far identified with these properties, J1010+2358 (Fig.~\ref{fig:mch}, bottom). For the latter, we consider separately the fraction of stars imprinted at a $>10\%$, $>50\%$, $>70\%$, $>80\%$ and $>90\%$ level by their massive PISN progenitor. In each case we assume that at least one star fits the criteria, J1010+2358 (Sec.~\ref{data-monoenrich}).

\begin{figure*}
\begin{center}
\includegraphics[width=0.7\hsize]{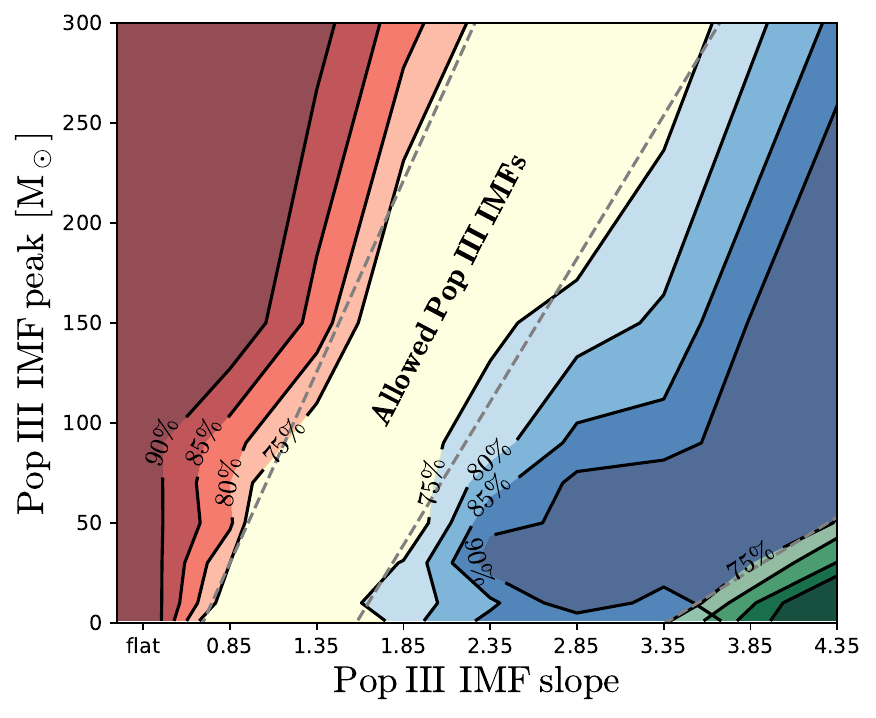} 
\caption{Confidence levels at which we can exclude a Pop~III IMF  with characteristic mass $m_{\rm ch}$ and slope $x$, based on: (i)~the non-detection of mono-enriched PISNe descendants in the SAGA catalogue at ${\rm [Fe/H]}<-2.5$ (top-left; red contours), and (ii)~the discovery of J1010+2358 among 15000 VMP LAMOST stars (bottom-right), assuming that J1010+2358 has inherited $>70\%$ (green contours) or $>90\%$ (blue contours) of its metals from a 250-260$\:{\rm M_\odot}$ PISN. Dashed lines show linear fits to the $75\%$ confidence levels. The central, light-yellow area represents Pop~III IMFs that remain possible.}
\label{fig: confidence}
\end{center}
\end{figure*} 

Fig.~\ref{fig:mch} shows our results when varying the characteristic mass of the Pop~III IMF (Eq.~\ref{e:IMF}) for a constant slope $x=2.35$ (left panels), and the slope $x$ for a characteristic mass $m_{\rm ch}=70\:{\rm M_\odot}$ (right panels). We find that a number of mono-enriched PISN descendants survive in the Galactic halo, spanning $\rm [Fe/H]<-5$ up to at least $\rm[Fe/H]=-2$ (Fig.~\ref{fig:mch}, top panels; see also \citealt{Magg2022}). In all cases, the predicted fractions show a maximum at ${\rm -5 \leq [Fe/H]} \leq-4.5$. At lower metallicities, the fraction decreases, since only the lowest mass PISNe ($m_{\rm PISN}<150\:{\rm M_\odot}$) produce such little iron to result in ${\rm [Fe/H]_{\rm ISM}<-5}$ \citep{salvadori2019probing}.
At $\rm[Fe/H]>-4.5$, the fraction of mono-enriched PISN descendants decreases strongly again, because of the gradual dominance of stars mainly imprinted by normal Pop~II SNe (Figs.~7-8 in \citealp{Koutsouridou2023}).

We find that the expected fraction of both partially and mono-enriched PISN descendants increases both with increasing $m_{\rm ch}$ and when the IMF gets flatter (decreasing $x$), naturally, since in these cases more PISNe are formed. From Fig.~2 (top), it seems like certain models are disfavoured, e.g. $m_{\rm ch} = 300 \:{\rm M_\odot}$, since they suggest that some mono-enriched PISN descendants should have already been identified at ${\rm [Fe/H]}<-2.5$.
However, to limit the Pop~III IMF it is necessary to quantify the confidence level of the exclusion and to explore the entire parameter space (Fig.~3).

In the bottom panels of Fig.~\ref{fig:mch}, we focus on the descendants of massive $250-260\:{\rm M_\odot}$ PISNe, and quantify their predicted fraction, for a given $f_{\rm PISN}$, with respect to the overall stellar population at ${\rm [Fe/H]}\leq-2$. Our predictions are compared to the unique star, J1010+2358, so far identified to be consistent with a minimum 10$\%$ enrichment by a massive PISN within the LAMOST VMP stellar sample (Sec.~\ref{data-monoenrich}). Since other stars partially imprinted by massive PISNe might be present in this sample, we cannot exclude IMFs that lie {\it above} the observational data point. Therefore, if future observations prove J1010+2358 to be enriched only at a $<50\%$ level by a massive PISN, we will not be able to put significant constraints on the Pop~III IMF. What if instead J1010+2358 has been predominantly ($>50\%$) enriched by a massive PISN?

For each characteristic mass $m_{ch}$, and Pop~III IMF slope $x$ (Eq.~\ref{e:IMF}), we compute the mean probability, $P_0$, of \textit{not} having detected any mono-enriched PISN descendant among the 962 SAGA stars with ${\rm[Fe/H]}<-2.5$, and its statistical error, $\delta P_0$, between 200 model realizations:
\begin{equation}
P_0 = \frac{ (N_{\rm tot}-N_{\rm desc})! (N_{\rm tot}-N_{\rm obs})! }{N_{\rm tot}!(N_{\rm tot}-N_{\rm desc}-N_{\rm obs})!}\approx \bigg(1-\frac{N_{\rm desc}}{N_{\rm tot}}\bigg)^{N_{\rm obs}},
\end{equation}
where $\rm N_{\rm tot}$, $\rm N_{\rm desc}$ and $\rm N_{\rm obs}$ are the total number of expected stars, PISN descendants, and observed stars in each [Fe/H] bin\footnote{The cumulative probability of {\it not} observing PISN descendants at ${\rm [Fe/H]}<-2.5$ is the product of the probabilities in each bin.} and the approximation is valid for $\rm N_{\rm tot}>>N_{\rm obs}$. The probability of finding {\it at least} one star enriched by a massive 250-260$\:{\rm M_\odot}$ PISN at a $>X\%$ level is given by $1-P_0 = 1-(1-\frac{N_{X\%}}{N_{tot}})^{N_{\rm obs}}$, using the above equation. If one of these two probabilities result $\rm P+\delta P<0.25$, then we can exclude the assumed Pop~III IMF with a confidence level of $[\rm 1-(P+\delta P)]>75\%$.

Fig.~\ref{fig: confidence} illustrates our results. By exploiting the {\it non-detection} of mono-enriched PISN descendants among the observed stars at ${\rm[Fe/H]}<-2.5$, we can already exclude a flat IMF at a $>90\%$ confidence level. At a $>75\%$ confidence level, we can also exclude Pop~III IMFs with characteristic mass $m_{\rm ch}/{\rm M_\odot}>191.16x-132.44$ (leftmost dashed line in Fig.~\ref{fig: confidence}). If J1010+2358 is found to be enriched at a $70\%$ level by a 250-260$\:{\rm M_\odot}$ PISN, we would only be able to exclude IMFs with $m_{\rm ch}/{\rm M_\odot}<53.33x-179.03$ (rightmost dashed line in Fig.~\ref{fig: confidence}). If, however, J1010+2358 proves to be enriched $>90\%$ by a massive PISN, it would provide much tighter constraints on the Pop~III IMF; excluding a Salpeter-like slope ($x=2.35$) with $m_{\rm ch}\lesssim130\:{\rm M_\odot}$, and Pop~III IMFs with steeper slopes at a confidence level that increases with increasing $x$ and decreasing $m_{\rm ch}$ (dashed line in Fig.~\ref{fig: confidence} corresponding to $m_{\rm ch}/{\rm M_\odot}=143.21x-225.94$ at a $75\%$ confidence level).

\section{Discussion and Conclusions}
\label{Discussion}

After decades of dedicated searches, the first very promising PISN descendant, J1010+2358, has been discovered within the LAMOST survey \citep{Xing2023}. Here, we have confirmed that a contribution from a massive (250-260$\:{\rm M_\odot}$) PISN is necessary to reproduce its abundance pattern. However, the PISN contribution can be as low as $10\%$ with the rest of its metals coming from other SNe (Fig.~\ref{fig:Xing_fit}). Exploiting our novel cosmological galaxy formation model of a MW-analogue, {\sc NEFERTITI} \citep{Koutsouridou2023}, we investigate the implications for the Pop~III IMF of: (i)~the non-detection of mono-enriched PISN descendants with ${\rm [Fe/H]}<-2.5$ in the Galactic halo, and (ii)~the discovery of J1010+2358 at $\rm[Fe/H]=-2.4$ (Fig.~\ref{fig:mch}). 

The non-detection of mono-enriched PISN descendants allows us to exclude at a $>90\%$ confidence level a flat Pop~III IMF (Fig.~\ref{fig: confidence}), which has both been favoured (e.g. \citealp{Hirano2017, Parsons2022}) and disfavoured (e.g. \citealp{deBen2017}) by previous studies. Furthermore, we can exclude at a $>75\%$ confidence level, a Larson type Pop~III IMF with characteristic mass $m_{\rm ch}/{\rm M_\odot} > 191.16x-132.44$, where $x$ is the slope (Eq.~\ref{e:IMF}). This area rules out the shallower IMFs considered for Pop~III stars (e.g. \citealp{Tarumi2020, Chen2022}). On the other hand, the constraining power of J1010+2358's discovery, depends on the fraction of its PISN metal enrichment. If $>90\%$ of J1010+2358's metals have been inherited from a massive PISN progenitor, then the parameter space of the Pop~III IMF can be significantly reduced; where $m_{\rm ch}/{\rm M_\odot} < 143.21x-225.94$ is excluded at a 75$\%$ confidence level. This rules out Salpeter-like IMFs with $m_{\rm ch}<130\:{\rm M_\odot}$ commonly adopted in literature (e.g. \citealp{Pallotini2015, Komiya2016, Trinca2023}). If, however, the PISN contribution is only $<70\%$, then the Pop~III IMF cannot be further constrained. 

Our results are inevitably subject to model assumptions and uncertainties, such as the instantaneous mixing approximation and the adopted stellar yields (see \citealp{Koutsouridou2023}). However, they represent a big leap forward in our understanding of the mass distribution of the first stars. 

To make further progress, we need to: (i)~assess the PISN enrichment level of J1010+2358 by measuring additional key elemental abundances such as C, Al and K (Fig.~\ref{fig:Xing_fit}); and (ii)~significantly increase the number of stars with high quality spectra. Indeed, here we have shown that even a single detection of a true PISN descendant can dramatically impact our view on the nature of the first stars.

\section*{Acknowledgements}
This project has received funding from the ERC Starting Grant NEFERTITI H2020/808240.



\end{document}